\documentclass[twocolumn,superscriptaddress,aps,preprintnumbers,amsmath,amssymb,prd]{revtex4-1}

\usepackage{graphicx}
\usepackage{epstopdf}
\usepackage{dcolumn}
\usepackage{mathtools}
\usepackage{bm}
\usepackage{hyperref}
\usepackage{ulem}
\usepackage{color}
\usepackage{amsmath}
\usepackage{comment}
\usepackage{siunitx}
\usepackage{here}

\def\Mpl{M_\mathrm{Pl}}
\usepackage{braket}

\newcommand{\bs}{\boldsymbol}
\newcommand{\tx}{\text}

\allowdisplaybreaks

\begin{document}

\title{
	Probing
	Axion-like Particles
	via CMB Polarization
}

\author{Tomohiro Fujita}
\email{tfujita@icrr.u-tokyo.ac.jp}
\affiliation{Institute for Cosmic Ray Research, The University of Tokyo, Kashiwa, 277-8582, Japan}

\author{Yuto Minami}
\email{yminami@rcnp.osaka-u.ac.jp}
\affiliation{Research Center for Nuclear Physics, Osaka University, Ibaraki, Osaka, 567-0047, Japan}
\author{Kai Murai}
\email{kmurai@icrr.u-tokyo.ac.jp}
\affiliation{Institute for Cosmic Ray Research, The University of Tokyo, Kashiwa, 277-8582, Japan}
\affiliation{Kavli Institute for the Physics and Mathematics of the Universe (WPI), The University of Tokyo, Kashiwa, 277-8583, Japan}

\author{Hiromasa Nakatsuka}
\email{hiromasa@icrr.u-tokyo.ac.jp}
\affiliation{Institute for Cosmic Ray Research, The University of Tokyo, Kashiwa, 277-8582, Japan}

\begin{abstract}
	Axion-like particles (ALPs) rotate the linear polarization of photons through the ALP-photon coupling and convert the cosmic microwave background (CMB) $E$-mode to the $B$-mode.
    We derive the relation between the ALP dynamics and the rotation angle by assuming that the ALP $\phi$ has a quadratic potential, $V=m^2\phi^2/2$.
    We compute the current and future sensitivities of CMB observations to the ALP-photon coupling $g$, 
    which can reach $g=4\times 10^{-21}\,\mathrm{GeV}^{-1}$ for $10^{-32}\,\mathrm{eV}\lesssim m\lesssim 10^{-28}\,\mathrm{eV}$ and extensively exceed the other searches for any mass $m\lesssim 10^{-25}\,\mathrm{eV}$.
    We find that the fluctuation of the ALP field at the observer, which has been neglected in previous studies, can induce significant isotropic rotation of the CMB polarization.
    The measurements of isotropic and anisotropic rotation allow us to put bounds on relevant quantities such as 
    the ALP mass $m$ and the ALP density parameter $\Omega_\phi$.
    In particular, if LiteBIRD detects anisotropic rotation, we obtain the lower bound on the tensor-to-scalar ratio as
    $r > 5 \times 10^{-9}$.
\end{abstract}

\maketitle

\section{Introduction}
\label{sec_introduction}

The axion has attracted much interest in particle physics and cosmology.
Peccei and Quinn originally introduced the QCD axion to solve the strong $CP$ problem~\cite{Peccei:1977hh}.
In recent decades, it has been found that string theory predicts a number of axion-like particles (ALPs), 
which have a broad range of masses and couplings to gauge fields. 
Such ALPs are expected to be ubiquitous in our universe, which provides a paradigm called the string axiverse~\cite{Arvanitaki:2009fg}.
ALPs are good candidates for dark matter~\cite{Preskill:1982cy,Abbott:1982af,Dine:1982ah}, 
and their ultralight mass may solve the so called small-scale crisis in cosmology~\cite{Weinberg:2013aya,Press:1989id,Sahni:1999qe,Hu:2000ke,Peebles:2000yy}.
Moreover, an extremely light ALP can be responsible for dark energy~\cite{Frieman:1995pm,Kim:2002tq,Tsujikawa:2013fta,Panda:2010uq,Ibe:2018ffn}.

Searches for axions and ALPs use various methods~\cite{Marsh:2015xka,Irastorza:2018dyq}.
In particular, the coupling to photon is one of the most promising detection schemes.
When photons travel through the ALP background, their polarization angle rotates, 
which is known as ``cosmic birefringence''~\cite{Carroll:1998zi,Lue:1998mq,Feng:2004mq,Feng:2006dp,Liu:2006uh}.
To measure the rotation angle, 
we need to observe some known polarized photons, e.g., astronomical targets~\cite{Alighieri:2010eu,Fujita:2018zaj,Ivanov:2018byi,Caputo:2019tms,Poddar:2020qft},
laser interferometers~\cite{Obata:2018vvr,Liu:2018icu,Nagano:2019rbw},
and the cosmic microwave background (CMB)~\cite{Agrawal:2019lkr,Fedderke:2019ajk,Sigl:2018fba}.
The CMB photons acquire uncorrelated $E$- and $B$-mode polarization
when they are emitted at the last scattering surface (LSS).
The cosmic birefringence mixes the $E$- and $B$-modes, 
which results in the $EB$ cross correlation~\cite{Lue:1998mq,Pospelov:2008gg,Finelli:2008jv,Lee:2013mqa,Zhao:2014yna,Lee:2016jym,Liu:2016dcg}.
Since the $EB$ cross correlation vanishes in parity-conserving models,
its detection is a smoking gun of parity-violating
phenomena, e.g., the ALP-photon coupling.

In this paper,
we formulate the polarization rotation of CMB photons induced by the ALP field 
with a small mass $m\lesssim 10^{-25}\,\mathrm{eV}$.
We take into account the fluctuation of the ALP at the observer in addition to the fluctuation at the LSS \cite{Pospelov:2008gg,Lee:2013mqa,Zhao:2014yna,Lee:2016jym} and background dynamics \cite{Lue:1998mq,Finelli:2008jv,Sigl:2018fba}.
We find that these three components of the ALP field independently contribute to spatially isotropic and anisotropic rotation of the CMB polarization.
We also discuss the fact that the oscillation of the ALP during the last scattering process suppresses the polarization rotation.
Since the ALP is a hypothetical particle and its abundance is bounded only from above,
we cannot put an upper constraint on the ALP-photon coupling from the observational upper bound on the CMB birefringence.
Instead, we derive the potential best sensitivity to the ALP-photon coupling by considering the largest allowed ALP abundance.
Then we show the current and future sensitivities of the CMB observations to the ALP-photon coupling against the ALP mass. 
Finally we find that we can put bounds on the ALP mass $m$,
the ALP density parameter $\Omega_\phi$, and the tensor-to-scalar ratio $r$ using the bounds from other experiments and the different ALP mass dependences of the sensitivities of isotropic and anisotropic rotation to ALP-photon coupling.

This paper is organized as follows.
In the section II, we formulate the relation between the ALP field and observational signals.
In the section III, we describe the dynamics of the ALP background and its perturbations, including its oscillation effect at the LSS.
The section IV shows our main result, the current and future sensitivities of the axion-photon coupling.
The section V is devoted to a summary and discussion.

\section{Birefringence by ALP}
\label{sec_Birefrengence}
Let us consider a Lagrangian with an ALP field $\phi$ coupled to photon:
\begin{equation}
\mathcal{L}=-\frac{1}{2}\partial^\mu \phi \partial_\mu \phi-V(\phi)
-\frac{1}{4}F_{\mu\nu}F^{\mu\nu} +\frac{1}{4}g\phi F_{\mu\nu}\tilde F^{\mu\nu},
\label{eq_lagrangian}
\end{equation}
where $V(\phi)$ is the ALP potential, 
$g$ is the coupling constant, $F_{\mu\nu}$ is the 
electromagnetic tensor, and $\tilde F^{\mu\nu}$
is its dual. 

When we observe CMB photons emitted at the LSS,
the difference in the ALP field values between the observer (``obs'') and the LSS rotates the polarization plane of the CMB photons by $\alpha= \left( \phi(x_\mathrm{obs})-\phi(x_\mathrm{LSS})\right)g/2$~\cite{Harari:1992ea}.
As $\phi$ depends on space-time,
let us decompose it
into the spatial-average term and the spatial fluctuation term as,
\begin{align}
\phi (t_\mathrm{LSS},d_\mathrm{LSS}\hat{\bs n}) &= \bar{\phi}_\mathrm{LSS} +  \delta\phi_\mathrm{LSS},
\\
\phi (t_\mathrm{obs},\mathbf{0} ) &= \bar{\phi}_\mathrm{obs} +  \delta\phi_\mathrm{obs},
\end{align}
where $\mathbf{0}$ is a position of the observer,
$\hat{\bs n}$ is the sky direction,
and $d_\mathrm{LSS}$ is the distance to the LSS.
Then, the direction-dependent rotation angle is given by
\begin{align}
    \alpha(\hat{\bs n}) = \frac{g}{2}\left( \Delta\bar{\phi} +\delta\phi_\mathrm{obs} - \delta\phi_\mathrm{LSS}\right),
    \label{eq:alpha vs g and phi}
\end{align}
with $\Delta \bar{\phi} \equiv \bar{\phi}_\mathrm{obs} - \bar{\phi}_\mathrm{LSS}$.

Here we elaborate on the decomposition of $\phi$ into $\bar\phi$ and $\delta\phi$ in Fourier space.
The contributions to $\phi(t,\bm x)$ from the Fourier modes with wavelengths longer than $d_\mathrm{ LSS}$ 
take the same value at the observer and the LSS position. 
On the other hand, if the wavelengths are shorter than $d_\mathrm{LSS}$, their contributions are uncorrelated,
because their phases are different at these distant positions.
Thus, we decompose 
$\phi$ in the following way:
\begin{align}
	\bar{\phi} (t) &\equiv
	\int_0^{k_*}
	\frac{\mathrm{ d}^3 k}{(2\pi)^3}\,
	e^{i\bm{ k}\cdot \bm{x}}\,
	\phi_{\bm k} (t),
\\
\delta\phi(t,{x}) &\equiv
    \int_{k_*}^\infty\frac{\mathrm{ d}^3 k}{(2\pi)^3}\, e^{i\bm{ k}\cdot {\bm x} }\phi_{\bm k}(t),
\end{align}
where $k_* = d_\mathrm{ LSS}^{-1}\approx H_0/3$ is the splitting scale,
$H_0 \equiv H(t_0)$ denotes the Hubble constant at $t_0$, 
and we set the present scale factor $a(t_0)=1$.
Here, $\delta{\phi}_\mathrm{obs}$ and $\delta{\phi}_\mathrm{LSS}$ are uncorrelated,
while $\bar{\phi}_\mathrm{obs}(t)$ and $\bar{\phi}_\mathrm{LSS}(t)$ 
follow the same background dynamics.
We will calculate $\delta\phi_\mathrm{ obs}$ and $\delta\phi_\mathrm{ LSS}$ separately as independent perturbations.
We assume that inflation generates
the Gaussian fluctuation of $\phi_{\bm k}$
characterized by
the power spectrum $\mathcal{P}_\phi(t, k)$,
$\langle \phi_{\bm k}(t) \phi_{\bm p}(t)\rangle
=(2\pi)^3\delta(\bm k-\bm p)\frac{2\pi^2}{k^3}\mathcal{P}_\phi(t,k).$

The spatial distribution of $\delta\phi_\mathrm{ LSS}$ results in the direction-dependent rotation angle $\alpha(\hat {\bs n})$, i.e. the anisotropic birefringence.
The sensitivity of the CMB observations to the anisotropic birefringence is often characterized by
$A_{\alpha} \equiv L(L+1)C_L^{\alpha\alpha}/(2\pi)$,
where $C_L^{\alpha\alpha}$ is the angular power spectrum of 
$\alpha(\hat{\bs n})$~\cite{Caldwell:2011pu}.
We assume that the scale-invariant power spectrum of the ALP field,
$\mathcal{ P}_\phi^{\mathrm{in}} = \left( {H_I} / (2\pi) \right)^2$,
is produced with the Hubble parameter $H_I$ during inflation.
Then we find a simple relation for $L\lesssim 100$~\cite{Caldwell:2011pu},
\begin{align}
    A_\alpha=
    \frac{g^2}{4}  \mathcal{ P}_\phi^{\mathrm{in}},
    \label{Aalpha}
\end{align}
where we ignore the ALP mass. 
Thus, the power spectrum of $\delta\phi_\mathrm{LSS}$ is measured through the
cosmic birefringence power spectrum.

On the other hand, any observations detect only a single realization of $\delta\phi_\mathrm{ obs}$ 
at the observer.
As the mean of the perturbation always vanishes $\langle\delta\phi\rangle=0$,
we evaluate the magnitude of $\delta\phi_\mathrm{ obs}$ by its variance $\langle \delta\phi_\mathrm{ obs}^2\rangle$.
Combined with the background contribution, 
the isotropic rotation angle $\bar \alpha$ is given by
\begin{equation}
    \bar\alpha=\frac{g}{2}\left(\Delta\bar\phi+\delta\phi_\mathrm{ obs}\right).
\label{alphabar}
\end{equation}

In summary, the cosmic birefringence by the ALP has the following three contributions:
(i) the ALP background motion $\Delta\bar\phi$, 
(ii) the anisotropic distribution of the ALP field 
at the LSS $\delta\phi_\mathrm{ LSS}$, and 
(iii) the ALP field fluctuation at the observer $\delta\phi_\mathrm{ obs}$.
$\Delta\bar\phi$ and $\delta\phi_\mathrm{ obs}$
generate the isotropic birefringence $\bar\alpha$, whereas $\delta\phi_\mathrm{ LSS}$ generates the anisotropic birefringence measured by $A_\alpha$.
Even when the ALP field does not change over time, $\Delta\bar\phi=0$,
the contributions (ii) and (iii) still remain because of the spatial fluctuations.

\section{ALP field dynamics}
\label{sec_Mass}

In this paper, henceforth, 
we employ the quadratic mass term as a simple model,
\begin{equation}
V(\phi)=\frac{1}{2} m^2 \phi^2. 
\label{eq:mass potential}
\end{equation}
In the spatially flat Friedmann-Lema{\^\i}tre-Robertson-Walker universe,
the metric perturbation is given by
$\mathrm{d}s^2 = a^2(\eta)\left[
-\mathrm{d}\eta^2 +(\delta_{ij} + h_{ij})\mathrm{d}x^i \mathrm{d}x^j
\right]$,
where we choose the synchronous gauge and $\eta$ is the conformal time.
Then the equations of motion (EoMs) for the background and the perturbation are obtained as~\cite{Zhao:2014yna}
\begin{align}
\bar{\phi}'' + 2\mathcal{H} \bar{\phi}' +a^2m^2 \bar{\phi} &= 0,
\label{eq:BG EoM}
\\
\delta\phi'' +2\mathcal{H}\delta\phi' -\nabla^2\delta\phi+a^2m^2\delta\phi 
	&= -\frac{1}{2}h'\bar{\phi}',
\label{eq:perturbation EoM}
\end{align}
where $h$ is the trace of $h_{ij}$.
Here, the source term, $-\frac{1}{2}h'\bar{\phi}'$, describes 
that the ALP perturbation is induced by the adiabatic perturbation
in proportion to the ALP background motion $\bar\phi'$~\cite{Caldwell:2011pu,Dave:2002mn}.
Nevertheless, we conservatively neglect the source term 
to estimate the robust contribution from $\delta\phi$ to birefringence,
which is inevitably generated by inflation irrespective of the background dynamics $\bar\phi(t)$.
Note that $\delta\phi_\mathrm{LSS}$ is not affected by the source term in any case,
because $\bar\phi$ does not evolve for $t<t_\mathrm{LSS}$ in the mass region of our interest.

We first solve the above EoMs for $\bar\phi$ and $\delta\phi_\mathrm{obs}$ in the Einstein-de Sitter (EdS) universe and discuss their behaviors analytically.
To determine the dynamics of $\bar \phi(t)$, one solves Eq.~\eqref{eq:BG EoM} and normally uses the initial field value to fix the integration constant.
Instead of using the initial field value, however, we employ the current density parameter of the ALP field, $\Omega_\phi\equiv
(\dot{\bar{\phi}}^2 + m^2\bar{\phi}^2)/6\Mpl^2 H_0^2$ to set the final field value.
We obtain simple expressions for $\Delta \bar\phi$ in two mass regions,
\begin{align}
    |\Delta\bar\phi|
	\simeq
	\begin{cases}
		\frac{2}{9}\sqrt{\frac{2}{3}\Omega_{\phi}}\, \frac{m}{H_0} \Mpl
		\qquad &(m\ll H_0)
	\\
		2\sqrt{\frac{2}{3}\Omega_{\phi}}\, \Mpl
		\qquad &(H_0 \ll m \ll H_\mathrm{ LSS})
	\end{cases},
	\label{light phi dif}
\end{align}
where $H_\mathrm{LSS}\equiv H(t_\mathrm{LSS})$.
In the upper row of Eq.~\eqref{light phi dif},
since the ALP mass is too small to start oscillating by now,
it is always in the slow-roll regime; thus $\bar\phi$ evolves only a little.
In the lower row, on the other hand,
the ALP began to oscillate during the epoch, $ t_\mathrm{LSS} \ll t \ll t_0$;
thus the ALP field value has been damped by Hubble friction,
which leads to $|\bar\phi_\mathrm{ obs}| \ll |\bar\phi_\mathrm{ LSS}|$.
As a result, $\Delta\bar\phi$ is suppressed by the factor $m/H_0$ in the lighter ALP case
compared to the intermediate-mass case with $H_0 \ll m \ll H_\mathrm{LSS}$.

The analytic solution for Eq.~\eqref{eq:perturbation EoM} in the
EdS universe also gives the approximate solution of $\delta\phi_\mathrm{obs}$.
Ignoring the mass and source terms, one finds the solution in Fourier space as
\begin{equation}
\phi_{\bm k} (\eta
)=  \frac{3\hat \phi_{\bm k}^\mathrm{ in}}{k^3\eta^3}\big[\sin(k\eta)-k\eta \cos(k\eta)\big],
\end{equation}
where we used the initial condition $\phi_{\bm k}(\eta_\mathrm{ in})=\hat \phi^\mathrm{ in}_{\bm k}$ and $\phi_{\bm k}'(\eta_\mathrm{ in})=0$
with a initial time $\eta_\mathrm{ in}$ which satisfies $k\eta_\mathrm{ in}\ll 1$ for relevant wave numbers.
The variance of $\delta\phi_\mathrm{obs}$ is computed as
\begin{equation}
    \langle \delta\phi_\mathrm{ obs}^2\rangle =
    \int^\infty_{k_*}\! \int^\infty_{k_*} \frac{\mathrm{ d}^3k\mathrm{ d}^3p}{(2\pi)^6} \langle \phi_{\bm k} (\eta_0) \phi_{\bm p} (\eta_0) \rangle
    \approx 0.9\, \mathcal{P}_\phi^\mathrm{ in} ,
\end{equation}
where $\eta_0$ is the present conformal time.
The above result implies that the typical size of $|\delta\phi_\mathrm{obs}|$ is $(\mathcal{P}_\phi^\mathrm{ in})^{1/2}$,
while its actual value in our universe is determined only in a stochastic manner due to cosmic variance.

Finally we consider the effective time average of the 
oscillating ALP field over
the thickness of the LSS
for $m\gtrsim H_\mathrm{LSS}$~
\cite{Capparelli:2019rtn}:
\begin{equation}
	\langle \bar{\phi} \rangle_{\mathrm{LSS}} = \int \mathrm{d}T\, g(T) \bar{\phi}\left(t(T)\right),
    \label{eq_LSSwashout}
\end{equation}
where $g(T)$ is the visibility function which describes 
the probability density that a CMB photon, observed now, scattered at the cosmic temperature $T$.
We approximate $g(T)$ by a Gaussian function,
\begin{equation}
	g(T) \simeq \frac{1}{\sqrt{2\pi}\sigma_T}\exp\left[ -\frac{(T-T_L)^2}{2\sigma_T^2} \right],
\end{equation}
where
$T_L = 2941\,\si{K}$ and $\sigma_T = 248\,\si{K}$ are the fitting parameters of the visibility function~\cite{Weinberg:2008zzc}.
We numerically find that $\langle \bar{\phi} \rangle_{\mathrm{LSS}}$ exponentially decays 
as $m$ increases for $m \gtrsim H_{\mathrm{LSS}}$. 
Note that, for an even heavier ALP, $|\langle \bar{\phi} \rangle_{\mathrm{LSS}}|$ becomes smaller than $|\bar{\phi}_\mathrm{obs}|$, and then $\bar{\phi}_\mathrm{obs}$ dominates $\Delta \bar{\phi}$.
We also apply the same
damping effect to $\delta\phi_{\mathrm{LSS}}$,
since the mass dependence of $\delta\phi_\mathrm{LSS}$ is effectively the same as the background $\bar{\phi}_\mathrm{LSS}$.

\section{Sensitivity of the CMB observation}
\label{sec_Results}

We numerically calculate the ALP field dynamics based on the $\Lambda$CDM model 
by solving Eqs.~\eqref{eq:BG EoM} and \eqref{eq:perturbation EoM} 
together with the Friedmann equation,
\begin{equation}
    H = H_0 \sqrt{\Omega_{\Lambda}+\Omega_M (a^{-3}+a_{\mathrm{eq}}a^{-4})},
\end{equation}
where $\Omega_M \simeq 0.31$ is the density parameter of matter and $a_{\mathrm{eq}} \simeq 1/3400$ is the scale factor at matter-radiation equality.
The upper bound of $\Omega_{\phi}$, which appeared in Eq.~\eqref{light phi dif}, is fixed as
\begin{align}
	\Omega_{\phi} \leq
	\begin{cases}
		\Omega_{\Lambda} 
		&(m \leq 9.26\times 10^{-34} \,\si{\eV})
		\\
		0.006h^{-2}
		&(10^{-32}\,\si{\eV} \leq m \leq 10^{-25.5}\,\si{\eV})
	\end{cases},
	\label{Omega_phi_max}
\end{align}
where $\Omega_{\Lambda} \simeq 0.69$ is the density parameter of dark energy~\cite{Aghanim:2018eyx}.
$\bar{\phi}$ can be responsible for all dark energy if its mass is sufficiently small, while its equation of state (EoS) parameter $w$ deviates from $-1$ as $m$ increases.
We obtain the condition that $\Omega_\phi$ can be $\Omega_\Lambda$ as $m \leq 9.26\times 10^{-34} \,\si{\eV}$ by requiring $w= ( \dot{\bar{\phi}}^2 - m^2 \bar{\phi}^2 ) / ( \dot{\bar{\phi}}^2 + m^2\bar{\phi}^2)$ to lie within 
$w_\Lambda(t_0) = -1.04 \pm 0.10$~~\cite{Aghanim:2018eyx}.
For $m \geq 10^{-32}\,\si{\eV}$, 
the ALP behaves as dark matter once it
starts oscillating at $H\simeq m$.
CMB and large-scale structure observations
constrain the ALP with such a transition of $w$
as $\Omega_{\phi} h^2 \leq 0.006$~\cite{Hlozek:2014lca}.
For the intermediate-mass region $\SI{9.26e-34}{\eV} < m < \SI{e-32}{\eV}$, we linearly connect these
upper limits on $\Omega_{\phi}$ in the $\log m$-$\log\Omega_\phi$ plane.
This treatment is compatible with the constraint given in Ref.~\cite{Hlozek:2014lca}.
Moreover, we set the initial power spectrum of the ALP perturbation by fixing the tensor-to-scalar ratio $r = 2H_I^2/(\pi M_{\mathrm{Pl}}^2 \mathcal{P}_{\zeta})$, where $\mathcal{P}_{\zeta}=2\times 10^{-9}$. 

Since $\Omega_\phi$ and $r$ are bounded from above,
we can put upper bounds on $|\Delta \bar{\phi}|$ and $\mathcal{P}_\phi^\mathrm{in}$ or, equivalently, $|\bar{\alpha}|$ and $A_\alpha$ for a fixed $g$.
By combining these upper bounds and the observational upper bounds on the CMB birefringence,
we obtain the maximum sensitivity to $g$.
Note that, we obtain not the constraint but the sensitivity,
since the existence of the ALP is not observationally confirmed and the CMB birefringence does not necessarily take place.

\begin{table}[tbp]
\centering
\caption{Current bounds and projected sensitivities to the polarization rotation parameters. 
} 
\label{tab:rotation constraints}
\begin{tabular}{c||c|c|c|c}
 & Current &  LiteBIRD & SO & \shortstack[c]{CMB-S4\\like} \\
\hline
$|\bar\alpha|$ ($^{\circ}$) & $<0.6$~\cite{Aghanim:2016fhp} & 0.1~\cite{MinamiKomatsu:2020} &
-&
-\\ 
$A_{\alpha} $($10^{-3}\deg^2$) & $<8.3 $~\cite{Bianchini:2020osu,Namikawa:2020ffr} & $4.0 $~\cite{Pogosian:2019jbt} & $0.55$~\cite{Pogosian:2019jbt} & $0.033$~\cite{Pogosian:2019jbt}  \\
\hline
\end{tabular}
\end{table}
Substituting the numerically obtained $\Delta\bar\phi$, $\delta \phi_\mathrm{obs}$, and
$\delta\phi_\mathrm{LSS}$ into Eqs.~\eqref{Aalpha} and \eqref{alphabar}, we obtain the precise predictions
of the CMB polarization rotation. 
For simplicity, we evaluate $\delta\phi_\mathrm{ obs}$ in Eq.~\eqref{alphabar} by its 
root mean square (RMS).
With these predictions, we can translate the sensitivities of the CMB observations to $\bar\alpha$ and $A_\alpha$
into the sensitivities to 
$g$.
In Table~\ref{tab:rotation constraints}, the current and projected sensitivities at $68\%$ confidence level to $\alpha$ and $A_\alpha$
of \textit{Planck}~\cite{Aghanim:2016fhp}, the South Pole Telescope~\cite{Bianchini:2020osu}, Atacama Cosmology Telescope~\cite{Namikawa:2020ffr}, LiteBIRD~\cite{Sugai:2707729,MinamiKomatsu:2020}, Simons Observatory (SO)~\cite{Ade:2018sbj,Bryan:2018mva}, and a CMB-S4-like mission~\cite{abazajian2019cmbs4,Pogosian:2019jbt} are summarized~\footnote{In this paper,
we only use LiteBIRD for the projected sensitivity to $|\bar{\alpha}|$,
because sensitivity of polarization rotation is degenerate with calibration uncertainties on artificial rotation of polarization sensitive detectors~\cite{Komatsu:2010fb}
and we can not find sensitivities including calibration uncertainties for the other projects.}.

\begin{figure*}[t]
\hspace{-7.5cm}(a)\hspace{9.0cm}(b)\\ \vspace{-0.2cm}
\centering
    \begin{minipage}[t]{.47\textwidth}
	\centering
	\includegraphics[width=\linewidth ]{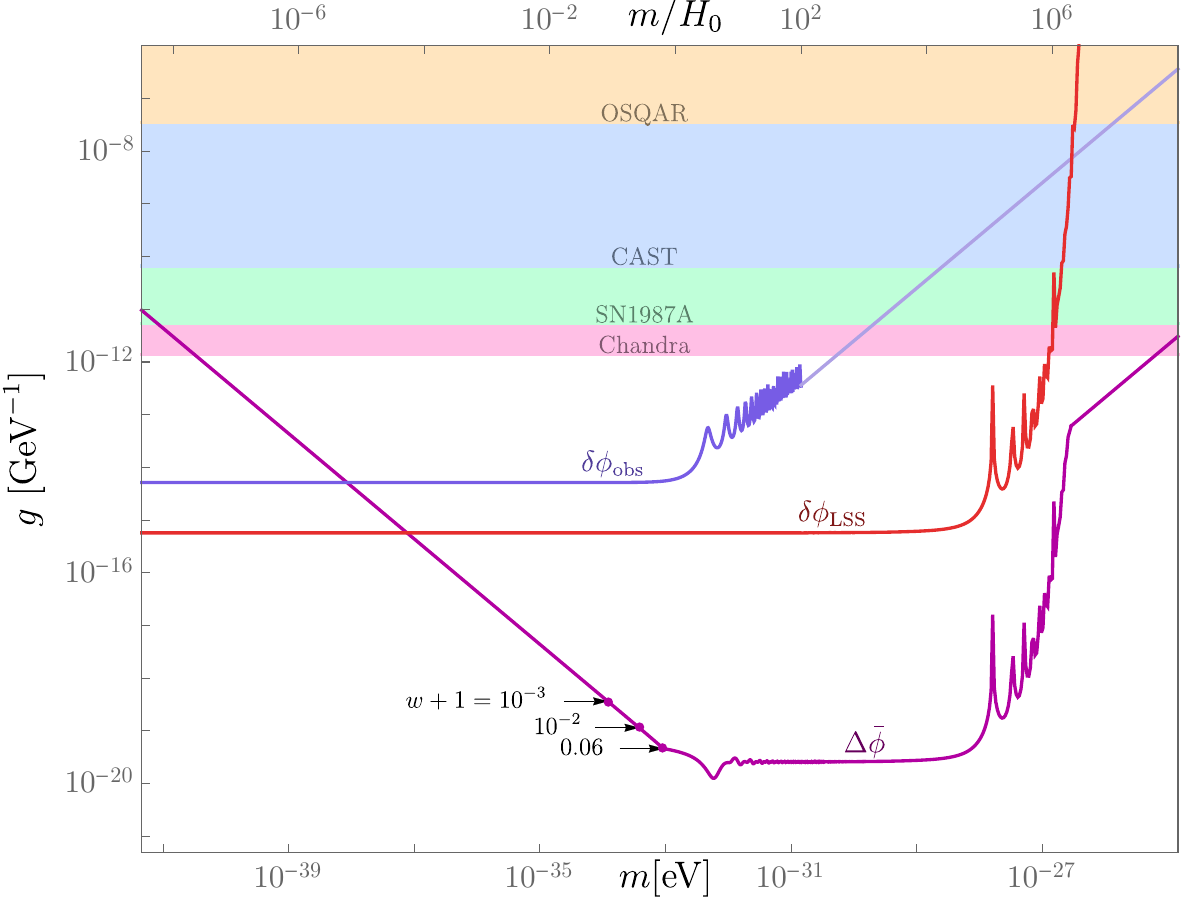}
	\label{fig_currentConst}
    \end{minipage}
	\hfill
    \begin{minipage}[t]{.47\textwidth}
	\centering
	\includegraphics[width=\linewidth ]{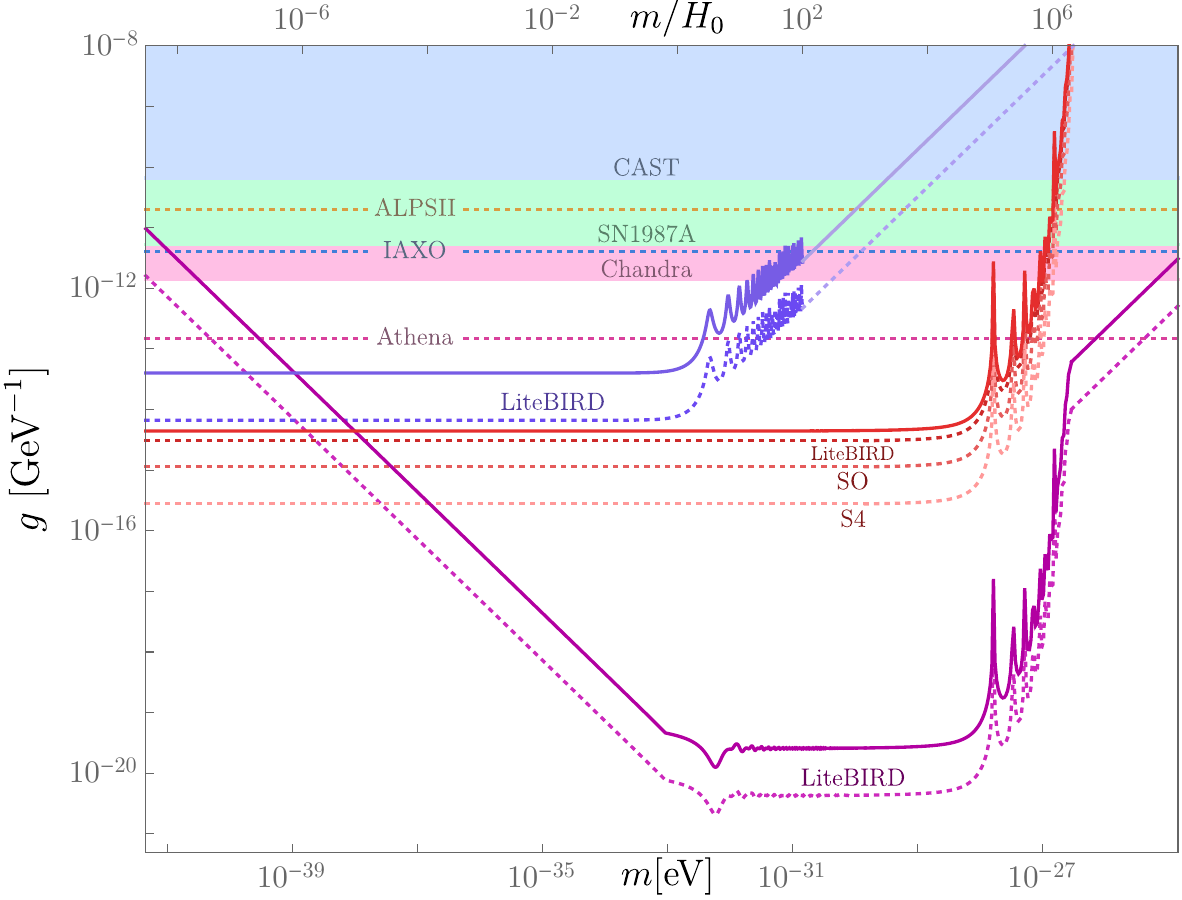}
	\label{fig_futureConst}
    \end{minipage}
    \caption{\label{fig:both}
        Panel~(a) shows the current best sensitivities to the ALP-photon coupling $g$ against the ALP mass $m$ from $\Delta \bar{\phi}$ (purple), $\delta \phi_{\mathrm{LSS}}$ (red), and $\delta \phi_{\mathrm{obs}}$ (blue).
        The purple line assumes the maximum $\Omega_\phi$ in Eq.~\eqref{Omega_phi_max}, and the others assume $r = 0.06$, while the sensitivities are lower for smaller $\Omega_\phi$ or $r$ as shown in Eqs.~\eqref{gDphi_light}-\eqref{eq_LSS sensitivity}.
		The purple dots show the current EoS parameter $w$ satisfying $w+1 = 10^{-3}, 10^{-2},$ and $0.06$ from left to right.
        We extrapolated the sensitivity from $\delta \phi_{\mathrm{obs}}$ with a light blue line proportional to $m$ for $m > 10^2 H_0$ due to high computational costs.
        The shaded regions have been excluded by
        OSQAR~\cite{Ballou:2015cka} (orange),
        CAST~\cite{Anastassopoulos:2017ftl} (light blue),
        SN1987A~\cite{Payez:2014xsa} (light green), and
        Chandra~\cite{Berg:2016ese} (pink).
        Panel (b) shows future sensitivities to the ALP-photon coupling $g$ of LiteBIRD,
        SO, and CMB-S4 (dotted lines), and current sensitivities (solid lines) 
		with $r = 10^{-3}$.
		The horizontal dotted lines show the projected sensitivities of
        ALPSII~\cite{Bahre:2013ywa} (orange),
        IAXO~\cite{Armengaud:2014gea,Irastorza:2018dyq} (light blue), and
        Athena~\cite{Conlon:2017ofb} (pink).
        }
\end{figure*}

Fig.~\ref{fig:both}(a) shows the current sensitivities to $g$ against $m$.
We obtain the best sensitivity owing to $\Delta\bar\phi$
by saturating Eq.~\eqref{Omega_phi_max}.
It is proportional to $m^{-1}$ for $m\lesssim 10^{-32}\,\si{\eV}$ and is flat
for a higher mass as explained in Eq.~\eqref{light phi dif}.
When $\Omega_\phi$ is smaller, the sensitivity to $g$ is reduced as $\Omega_\phi^{1/2}$.
The red line denotes the sensitivity originating from 
$\delta\phi_\tx{LSS}$. 
For $m\gtrsim \SI{e-28}{\eV}$, these two
lines exponentially blow up,
for the reason discussed around Eq.~\eqref{eq_LSSwashout}.
Note that the purple line increases in proportion to $m$ for $m\gtrsim \SI{e-27}{\eV}$, 
because $\bar\phi_\mathrm{obs}$ dominates $\Delta\bar{\phi}$ there.
The blue line represents the sensitivity contributed by $\delta\phi_\mathrm{obs}$,
which oscillates and attenuates for $H_0\lesssim m$.

We confirm that the parameter dependence of these sensitivities matches the analytic expressions obtained in the EdS universe.
Using the numerically computed coefficients, we find that 
the sensitivities to $g_X$, $g$ from $X\in\{\Delta\bar\phi, \delta\phi_\mathrm{LSS}, \delta\phi_\mathrm{obs}\}$, have the following expressions: 
\begin{align}
&g_{\Delta\bar\phi}(m\lesssim H_0) =
        3.0 \times \SI{e-18}{\per\GeV}\notag\\
        &\qquad\qquad \times\left(
        \frac{ |\bar\alpha| }{ \SI{0.6}{\degree}}
        \right)
        \left( \frac{\Omega_{\phi}}{\Omega_{\Lambda 0}} \right)^{-\tfrac{1}{2}}
        \left( \frac{m/H_0}{10^{-2}} \right)^{-1},
        \label{gDphi_light}
        \\
&g_{\Delta\bar\phi}(H_0 \lesssim m \lesssim H_{\mathrm{LSS}}) =
        2.6 \times \SI{e-20}{\per\GeV}\notag\\
        &\qquad\qquad\times\left( 
        \frac{ |\bar\alpha| }{0.6^{\circ} } \right)
        \left( \frac{\Omega_{\phi}h^2}{0.006} \right)^{-\tfrac{1}{2}},
        \label{gDphi_heavy}    
    \\
&g_{\delta\phi_\mathrm{obs}} ( m \lesssim H_0)
  =
  4.0
  \times 10^{-14}\,\si{\GeV}^{-1}
  \notag\\
  &\qquad\qquad\times
  \left(
  \frac{ |\bar\alpha| }{0.6^{\circ} } 
  \right)
  \left( \frac{r}{10^{-3}} \right)^{-1/2},
  \label{gdobs}
\\
  &g_{\delta\phi_{\small \mathrm{LSS}}} (m \lesssim H_{\mathrm{LSS}})
  =
  4.4 \times 10^{-15}\,\si{\GeV}^{-1}
\notag\\  
    &\qquad\qquad\times\left( \frac{ A_{\alpha }}{8.3 \times 10^{-3}\,\mathrm{deg}^2} \right)^{1/2}
  \left( \frac{r}{10^{-3}} \right)^{-1/2}.
  \label{eq_LSS sensitivity}    
\end{align}
Note that we separately describe the contribution to $g$ from each component in Eqs.~\eqref{gDphi_light}-\eqref{gdobs} to see its behavior, although the contributions from $\Delta\bar\phi$ and $\delta\phi_\mathrm{obs}$ to the observed $\bar\alpha$ are degenerate.

Forthcoming CMB observations will improve the sensitivity as shown in Fig.~\ref{fig:both}(b).
In Fig.~\ref{fig:both}, we see that the CMB observations of
$\bar\alpha$ and $A_\alpha$ can achieve considerably better sensitivities to $g$ than the existing and upcoming 
axion searches.
If a light ALP exists,
inflation automatically generates its fluctuations, and we can probe
$g$ through $\delta\phi_\mathrm{LSS}$ and $\delta\phi_\mathrm{obs}$.
If we additionally assume that the background ALP field has a significant energy density, $\Delta \bar{\phi}$ also contributes to $\bar{\alpha}$,
which provides another channel to probe $g$. In particular, when $\Delta\bar{\phi}$ is the maximum allowed value based on Eq.~\ref{Omega_phi_max},
the sensitivity of this channel can be $3\times 10^8$ times better than the current constraint in the near future.

It is interesting to notice that the detection of $A_\alpha$ puts a lower bound on $r$.
Since the ALP-photon coupling constant has the present upper bound, $g< 1.4 \times 10^{-12}\,\si{\GeV}^{-1}$,
once the observation fixes $A_\alpha$, we obtain
\begin{equation}
    r > 5 \times 10^{-9}
    \left( \frac{ A_{\alpha }}{4 \times 10^{-3}\,\mathrm{deg}^2} \right),
    \label{r lowbound}
\end{equation}
where we used Eq.~\eqref{eq_LSS sensitivity}.
If Athena improves the upper bound on $g$ by a factor of $10$, for instance, 
the lower bound on $r$ would increase by a factor of $100$.
Therefore, the observation of $A_\alpha$ combined with axion search experiments can give a lower bound on $r$, which is complementary to the CMB $B$-mode observation limiting $r$ from above.

We can explore the further implications of the possible detection of either $\bar\alpha$ or $A_\alpha$.
If $A_\alpha$ is detected, since $A_\alpha$ is contributed by $\delta\phi_\mathrm{LSS}$, it implies that non-negligible $r$ generated $\delta\phi_\mathrm{LSS}$.
At the same time, $r$ also generates $\delta\phi_\mathrm{obs}$, which induces $\bar{\alpha}$.
Combining Eqs.~\eqref{gdobs} and \eqref{eq_LSS sensitivity}, we expect $\bar{\alpha}$ contributed by $\delta\phi_\mathrm{obs}$ would be
\begin{equation}
    |\bar\alpha|
    \simeq 
    0.05^{\circ} \left( \frac{ A_{\alpha }}{4 \times 10^{-3}\,\mathrm{deg}^2} \right) ^{1/2}.
    \label{alphaAalpha}
\end{equation}
However, if $\bar\alpha$ is not observed well below the above expected value, 
we can constrain the ALP mass, $m\gg H_0$, because the ALP must oscillate by the present time to attenuate the contribution from $\delta\phi_\mathrm{obs}$.
At the same time,
we obtain the upper bound on $\Omega_\phi$,
by equating Eq.~\eqref{gDphi_heavy} to Eq.~\eqref{eq_LSS sensitivity},
\begin{equation}
    \Omega_\phi h^2 < 1.8\times 10^{-13} \left(\frac{ |\bar\alpha| }{0.05^{\circ}}\right)^2
    \left( \frac{ A_{\alpha }}{4 \times 10^{-3}\,\mathrm{deg}^2} \right)^{-1}
    \left( \frac{r}{0.06} \right),
    \label{eq_Omegaphi_up_biref}
\end{equation}
where not only $\bar\alpha$ but also $r$ is bounded from above, $r<0.06$~\cite{Akrami:2018odb,Ade:2018gkx}.

If $\bar\alpha$ is detected and the corresponding $A_\alpha$ given by Eq.~\eqref{alphaAalpha} is not observed,
we can conclude that the origin of the detected $\bar{\alpha}$ is $\Delta \bar{\phi}$.
Substituting the detected $|\bar\alpha|$ and the experimental upper bound on $g$ into $g_{\Delta\bar\phi}$, 
and using Eq.~\eqref{Omega_phi_max}, we obtain the allowed range of $m$ as 
\begin{equation}
10^{-8} \left(\frac{|\bar\alpha|}{0.3^{\circ}}\right) 
< \frac{m}{H_0} \lesssim
10^8 \left(\frac{|\bar\alpha|}{0.3^{\circ}}\right)^{-1},
\label{allowed mass range}
\end{equation}
where we used $\Delta\bar\phi\simeq \bar\phi_\mathrm{obs}= \sqrt{6\Omega_{\phi}}M_{\mathrm{Pl}}H_0/m$
for $m>3\times10^{-27}$ eV and Eq.~\eqref{gDphi_light} to derive the upper and lower bound, respectively.
Of course, we can develop similar arguments when both $\bar\alpha$ and $A_\alpha$ are detected.

\section{Summary and Discussion}
\label{sec_Conslusion}

In this paper, we have investigated the cosmic birefringence of CMB photons as a probe of the ALP-photon coupling $g$ under the assumption that the ALP has a quadratic potential with an extremely light mass $m \lesssim t_{\mathrm{LSS}}^{-1}$.
Using Eq.~\eqref{eq:alpha vs g and phi},
one can relate the observations of the birefringence angle to $g$ by calculating the ALP dynamics.

The background dynamics, $\Delta\bar{\phi}$, and the fluctuation at the observer, $\delta\phi_\mathrm{obs}$, induce isotropic birefringence,
while the fluctuation at the LSS, $\delta\phi_\mathrm{LSS}$, induces anisotropic birefringence.
The isotropic rotation induced by $\Delta \bar{\phi}$ largely depends on the energy fraction of the ALP and, with the maximum allowed energy fraction,
the sensitivity to $g$ extensively exceeds the current limits from other observations.

The same signal may be used to search for a quintessence field with a tiny $w+1$.
Even if the energy fraction of the ALP is negligible, the contributions of $\delta \phi_{\mathrm{obs}}$ and $\delta\phi_{\mathrm{LSS}}$ persist as long as the light ALP exists, and they have better sensitivities to $g$ than other observations depending on $r$.
In particular, we have found that $\delta \phi_{\mathrm{obs}}$
stochastically breaks parity symmetry and contributes to the isotropic rotation of the CMB polarization.
If the ALP is heavy enough to oscillate during last scattering, however,
the birefringence effect is drastically suppressed. 

If isotropic or anisotropic birefringence is observed by any CMB observations,
we can limit relevant parameters such as $r$, $\Omega_{\phi}$, and $m$.
Our findings derived in Eqs.~\eqref{r lowbound}-\eqref{allowed mass range}
can be summarized as follows.
(i) Eq.~\eqref{r lowbound}: once the anisotropic birefringence is detected, $r$ is bounded from below,
(ii) Eq.~\eqref{alphaAalpha}: if the anisotropic birefringence is detected, the isotropic one should also be observed; otherwise, the ALP is heavy, $m \gg H_0$.
(iii) Eq.~\eqref{eq_Omegaphi_up_biref}: if the anisotropic birefringence is detected but the isotropic one is not, we obtain the upper bound on the background ALP energy.
(iv) Eq.~\eqref{allowed mass range}: if the isotropic birefringence is detected but the anisotropic one is not, the ALP mass is bounded from both above and below.
These limits will be substantially improved with upcoming X-ray observations~\cite{Conlon:2017ofb} and future CMB observations~\cite{Sugai:2707729,Ade:2018sbj,abazajian2019cmbs4}.

We also comment on possible extensions of our work.
The quadratic mass potential Eq.~\eqref{eq:mass potential} used in our analysis should be seen as a toy model. 
For $\Omega_\phi\simeq\Omega_\Lambda$, this potential
requires the ALP field value $\bar\phi$ to be much larger than the Planck scale,
which may not be favored by UV completion. 
However, it is straightforward to compute
$\bar\alpha$ and $A_\alpha$ for the other ALP potential forms by following our procedure.

Although we have evaluated $\delta\phi_\mathrm{obs}$ using its RMS, 
$\delta\phi_\mathrm{obs}$ in our universe may deviate from the RMS value by chance. 
A dedicated statistical treatment is needed for more precise predictions. 
We have ignored the source term in Eq.~\eqref{eq:perturbation EoM},
while it might amplify $\delta\phi_\mathrm{obs}$.
We leave these intriguing problems for future work.

\begin{acknowledgments}
We would like to thank Ricardo Z. Ferreira, Masahiro Ibe, Masahiro Kawasaki, Eiichiro Komatsu, Ippei Obata, So Okano, G\"{u}nter Sigl, and Pranjal Trivedi
for fruitful discussions and productive comments.
K.M. was supported by World Premier International Research Center Initiative (WPI Initiative), MEXT, Japan and the Program of Excellence in Photon Science.
H.N. was supported by Advanced Leading Graduate Course for Photon Science.
This work was supported in part by the Japan Society for the Promotion of Science (JSPS) KAKENHI, Grant Number JP18K13537, JP19J21974, JP20K14497, and JP20J20248.
\end{acknowledgments}

\small
\bibliographystyle{apsrev4-1}
\bibliography{Ref}

\end{document}